\documentclass{PoS}

\usepackage{bbm} %maths

\newcommand{\be}{\begin{equation}}
\newcommand{\ee}{\end{equation}}
\newcommand{\bea}{\begin{eqnarray}}
\newcommand{\eea}{\end{eqnarray}}

\newcommand{\bi}{\begin{itemize}}
\newcommand{\ei}{\end{itemize}}

\boldmath
\title{On the spectral density of the Wilson operator}
\unboldmath

\ShortTitle{Spectral density}

\author{

\vspace*{-10mm}

\begin{flushright}
CERN-PH-TH-2011-196\\
HU-EP-11/39\\
\end{flushright}
\vspace*{7mm}

S.~Necco\\
CERN, Physics Departement, 1211 Geneva 23, Switzerland\\
E-mail: \email{necco@mail.cern.ch}}
\author{\speaker{A.~Shindler}\\
Institut f\"ur Elementarteilchenphysik, Fachbereich Physik,
\\ Humbolt Universit\"at zu Berlin, D-12489, Berlin, Germany\\
E-mail: \email{andrea.shindler@physik.hu-berlin.de}}

\abstract{We summarize our recent determination~\cite{Necco:2011vx} of the 
spectral density of the Wilson operator in the $p$-regime of Wilson chiral perturbation theory. 
We discuss the range of validity of our formula and a possible
extension to our computation in order to better understand the behaviour of the spectral 
density in a finite volume close to the threshold.}

\FullConference{The XXIX International Symposium on Lattice Field
  Theory - LAT2011\\ July 10-16 2011\\ the Village, Squaw Valley, Lake Tahoe, California, USA}

\begin{document}

\section{Introduction}
A theoretical description for the spectral density of the Wilson-Dirac operator is important mainly 
for two reasons. 
It has been recently shown in~\cite{Giusti:2008vb} that spectral observables computed
with Wilson fermions, such as the spectral density of the Wilson operator, can be a powerful 
tool to determine interesting quantities as the chiral condensate. Additionally, a theoretical understanding
of the behaviour of the spectral density close to the threshold might help in estimating stability 
bounds for HMC-like algorithms~\cite{DelDebbio:2005qa}.
Cutoff effects and finite size effects (FSE) modify the functional form of the spectral density 
of the Dirac operator, and Wilson chiral perturbation theory (W$\chi$PT) is the right tool 
to analyze these aspects.

In continuum chiral perturbation theory ($\chi$PT) in order to compute the spectral density $\rho_D(\gamma,m)$ of the 
Dirac operator $D_m = D + m$ with eigenvalues $\gamma_k + m$ one adds
a valence quark with mass $m_v$; the discontinuity  of the valence scalar quark condensate
along the imaginary axis 
of the $m_v$ plane 
is proportional to the spectral density~\cite{Chandrasekharan:1995gt,Osborn:1998qb}.
The valence quark condensate in partially quenched chiral perturbation theory (PQ$\chi$PT) can be computed 
using the graded group method of~\cite{Bernard:1993sv} or the replica method of~\cite{Damgaard:2000gh}. 
From the form of the spectral density in the continuum~\cite{Smilga:1993in,Osborn:1998qb,Giusti:2008vb} for $N_f=2$
\begin{eqnarray}
[\rho_D(\gamma,m)]_{NLO} &=& \frac{\Sigma}{\pi}\Bigg\{1+  \frac{\Sigma}{(4\pi)^2F^4}\Bigg[-\pi|\gamma| + 
m(3\bar{L}_6-1) +\nonumber\\
& +& 2\gamma\arctan \frac{\gamma}{m}-  2m\ln\left(\frac{\Sigma\sqrt{\gamma^2+m^2}}{F^2\mu^2} \right) -
m\ln\left(\frac{2\Sigma\gamma}{F^2\mu^2}   \right)\Bigg]\Bigg\},
\label{eq:rho_cont}
\end{eqnarray}
one can already observe that the spectral density is a very good candidate to compute the 
low energy constant $\Sigma$, because the leading order (LO) expression of $\rho_D(\gamma,m)$ 
is directly related to $\Sigma$ and the next-to-leading order (NLO) corrections vanish in the chiral limit.

\section{Spectral density of the Hermitean Wilson operator}

For the Wilson operator $D_W$ it is convenient to study the Hermitean operator 
$Q_m = \gamma_5 \left(D_W +m \right)=Q_m^{\dagger}$ with real eigenvalues $\lambda_k$. Giusti and L\"uscher
~\cite{Giusti:2008vb} have shown that the spectral density $\rho_Q$ of $Q_m$ renormalises multiplicatively as
\be
\left[\rho_Q\right]_R(\lambda, m_R)= Z_P\rho_Q(Z_P \lambda,m)\, , \qquad m_R = Z_m(m_0 - m_{\rm cr}) \, ,
\ee
where $Z_P$ is the renormalisation constant of the pseudoscalar density.
Moreover, once the action is improved the remaining cutoff effects are of O($am$) and they can
be removed by suitable improvement coefficients. It is also useful to determine $\Sigma$ to define the mode number
\be
N(\Lambda_1,\Lambda_2,m)= V\int_{\Lambda_1}^{\Lambda_2} 
d\lambda~ \rho_Q(\lambda,m),\;\;\;\; \Lambda_2\geq \Lambda_1\geq m
\label{eq:mn}
\ee
which is a renormalisation group invariant (RGI) quantity~\cite{Giusti:2008vb}.
To compute the spectral density $\rho_Q$ one can introduce a doublet of valence twisted mass
fermions $\chi_v$ with untwisted mass $m_v=m$ and twisted mass $\mu_v$; the spectral 
density is now related to the discontinuity along the imaginary axis of the twisted mass plane 
of the valence pseudoscalar condensate~\cite{Sharpe:2006ia}.
To compare the results obtained in W$\chi$PT for $\rho_Q$ with the continuum formula for $\rho_D$
we recall that the two spectral densities in the continuum are connected by the relation 
$\rho_Q(\lambda,m) = \frac{\lambda}{\sqrt{\lambda^2-m^2}}\rho_D\left(\sqrt{\lambda^2-m^2},m \right)$.

For all the details about our computation we refer to ref.~\cite{Necco:2011vx}. Here we simply recall that
for the valence and sea quark masses we have chosen to work in the $p$-regime, and for the lattice spacing
we have chosen to be in the generic small masses (GSM) regime where both masses are counted as O($a$), viz.
\be
m,\mu_v,a\sim O(p^2),\;\;\;1/L,1/T \sim  O(p)\, .
\ee
The final result~\cite{Necco:2011vx} for the spectral
density of the Hermitean Wilson operator in infinite volume for $N_f=2$ in terms if the PCAC quark mass is given by
\begin{eqnarray}
 [\rho_Q(\lambda,m_{\rm PCAC})+\rho_Q(-\lambda,m_{\rm PCAC})]_{NLO} &=& 2\left[\rho_Q(\lambda,m_{\rm PCAC})\right]_{NLO,cont} \nonumber \\ 
&+& \frac{2\Sigma\lambda}{\pi\sqrt{\lambda^2-m_{\rm PCAC}^2}}\Bigg[\frac{m_{\rm PCAC}^2\Delta}{\lambda^2-m_{\rm PCAC}^2}+\frac{16\hat{a}}{F^2}W_6\Bigg]\,,
\label{eq:rho_Q}
\end{eqnarray}
with $\hat{a}=2W_0a$. There are two types of corrections to the continuum formula: an overall shift of O($a$) parametrized by the W$\chi$PT LEC
$W_6$ and a modification of the shape of the spectral density with contributions of O($a$) and O($a^2$) 
parametrized by $\Delta=-\frac{16\hat{a}}{F^2} \left(\frac{W_8}{2} +\frac{W_{10}}{4} +\frac{\hat{a}W_8'}{M^2_{ss}}\right)$. 
It is important to notice that if we perform a non-perturbative improvement of the lattice theory 
the $W_6$ and $W_8$ terms vanish, and if we improve the axial current in the determination of the PCAC mass,
the $W_{10}$ term vanishes. We are thus left with remaining O($a^2$) cutoff effects proportional
to $W_8'$ that go to zero in the chiral limit.

\section{Comparison with numerical data and range of validity}
\begin{figure}[tb]
\includegraphics[width=0.45\textwidth,angle=0]{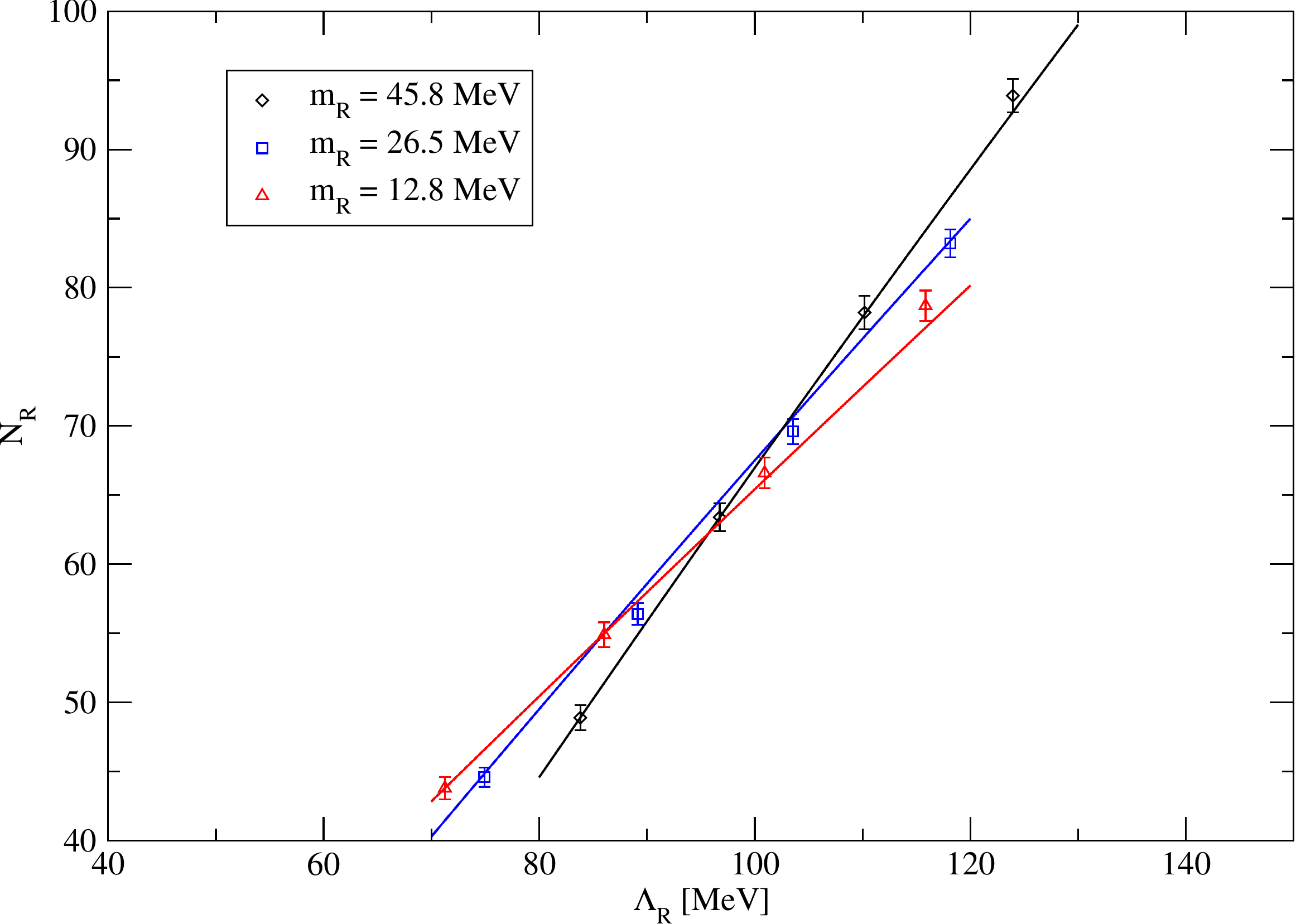}\hspace{1.0cm}
\includegraphics[width=0.45\textwidth,angle=0]{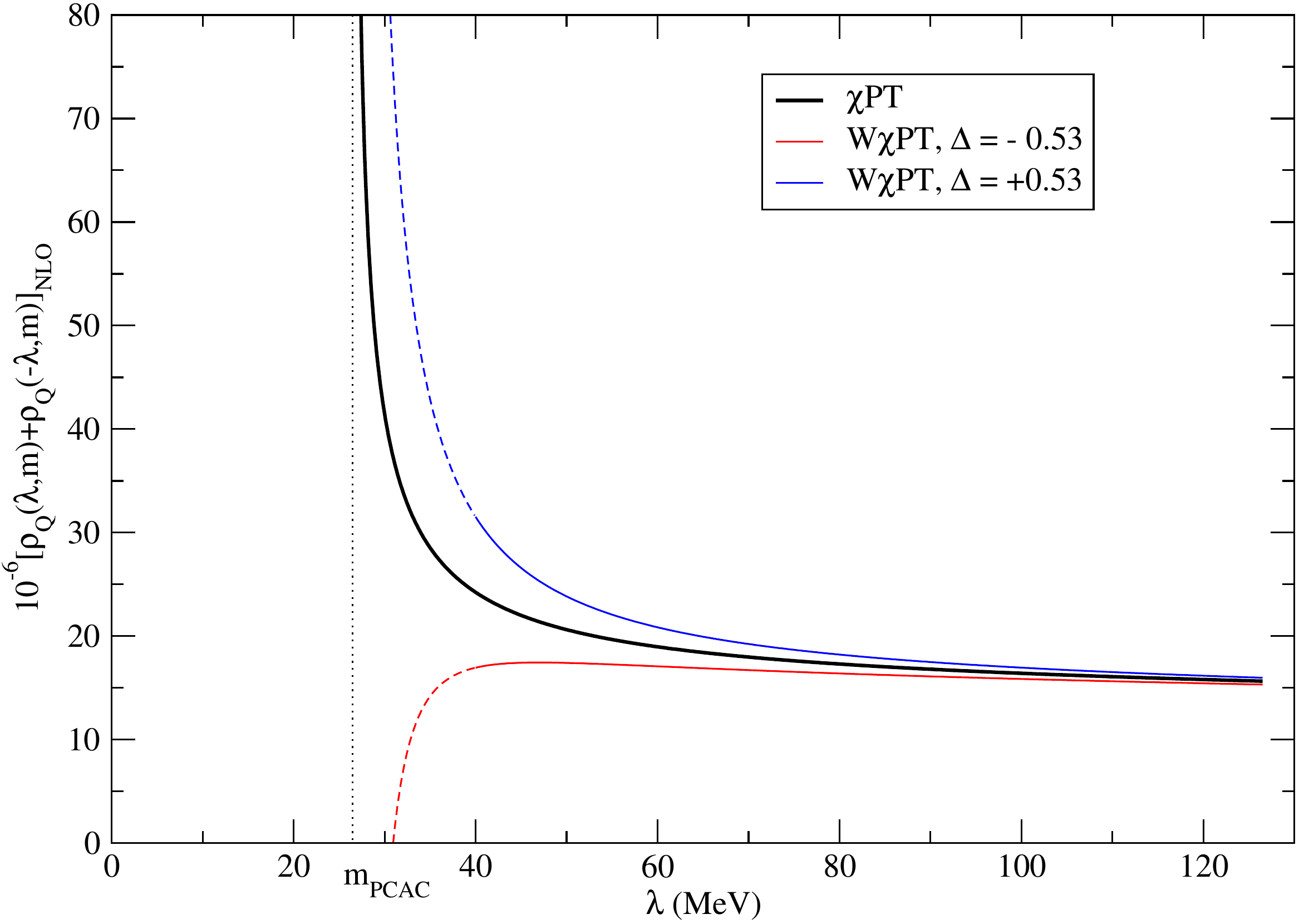}
\caption{Left plot: result of the global fit of the data published in~\cite{Giusti:2008vb} with our formula for the 
renormalised mode number (eqs.~\protect\ref{eq:mn} and~\protect\ref{eq:rho_Q}). The fit parameters are $\Sigma$, 
$\Delta$ and $\overline{L}_6$ and we obtain
$\chi^2/{\rm dof} = 0.91$. Right plot: 
the spectral density $[\rho_Q(\lambda,m)+\rho_Q(-\lambda,m)]_{NLO}$ in the infinite volume. 
We used the parameters $\Sigma=(275\;{\rm MeV})^3$, $m_{\rm PCAC}=26.5$ MeV, $F=90$ MeV, $\bar{L}_6=5$, $\mu=139.6$ MeV. 
The solid black line corresponds to the continuum $\chi$PT prediction, while 
the red (blue) lines correspond to the lattice W$\chi$PT prediction ($O(a)$-improved) 
on eq.~\protect\ref{eq:rho_Q} with $\hat{a}^2W'_8= \pm 5\cdot 10^6\; {\rm MeV}^4$, 
corresponding to $\Delta=\mp 0.53$.}
\label{fig:global_fit}
\end{figure}

To test our formula for the mode number (cfr. eqs.~\ref{eq:mn} and ~\ref{eq:rho_Q})
we compared it with the numerical data published in~\cite{Giusti:2008vb}.
The simulations of ref.~\cite{Giusti:2008vb} were performed with $N_f=2$ non-perturbative 
O($a$) improved Wilson fermions at lattice spacing $a\simeq 0.08$ fm and physical volume 
$L\simeq 2.5$ fm with $T=2L$. We have fixed $F=90$ MeV and the renormalisation scale 
$\mu=m_\pi$; we have performed a global fit at all the $3$ masses available and all the values 
of $\Lambda_R$ with $3$ fit parameters: $\Sigma$, $\Delta$ and $\overline{L}_6$.\footnote{In this proceedings with $\Sigma$ we denote its value
renormalised in the $\overline{MS}$ scheme at a scale of $2$ GeV.}
From the global fit we obtain
\be
\Sigma^{1/3} = 266(7) \, {\rm MeV}\,, \qquad \Delta = - 0.62(80)\,, \qquad \overline{L}_6 = 6(1)\,.
\ee 
The numerical data and our global fit are shown in the left plot of fig.~\ref{fig:global_fit}.
For $\Sigma$ we obtain a perfectly consistent result with Giusti and L\"uscher~\cite{Giusti:2008vb} without
performing any chiral extrapolation. We observe that we are also sensitive for a determination of 
$\overline{L}_6$, while for $\Delta$ we obtain a value consistent with zero.
We have also performed a fit fixing $F=80$ MeV obtaining compatible results within errors.

In our our paper~\cite{Necco:2011vx} we have also compared our result with the continuum $\chi$PT formula and with numerical 
data on a different range of $\Lambda$.
Even if the analysis presented in ref.~\cite{Necco:2011vx} is rather qualitative and more numerical data are needed,
our conclusion is that W$\chi$PT describes the numerical data 
better than continuum $\chi$PT sufficiently away from the threshold, even if we can't exclude that adding NNLO terms 
in the continuum formula could lead to a comparable improvement.
In the right plot of fig.~\ref{fig:global_fit} we show the NLO prediction for the spectral density in continuum $\chi$PT 
(cfr. eq. \ref{eq:rho_cont}, black curve) and W$\chi$PT (cfr. eq.~\ref{eq:rho_Q}) for two opposite
values of $\Delta$ (red curve $\Delta <0$ and blue curve $\Delta >0$). 
This plot is useful to make few observations. First we notice that in the region of $\lambda$ we have used 
to perform the global fit of fig.~\ref{fig:global_fit}, i.e. 
with $80 ~{\textrm{MeV}}~\lesssim ~\lambda~\lesssim 120$ MeV, and for a reasonable choice of parameters, cutoff effects are very small.
This is consistent with the fact that in the global fit we are not sensitive to determine $\Delta$, that turns out to be
consistent with zero. Cutoff effects become more visible in the range 
$40 ~{\textrm{MeV}}~\lesssim ~\lambda~\lesssim 60$ MeV and this might explain why all our fits 
in this range (cfr. figs. 3 and 4 of ref.~\cite{Necco:2011vx}) prefer a negative value for $\Delta$. In fact for $\Delta < 0$ 
we observe that over the whole range of $\lambda \gtrsim 40$ MeV the spectral density in W$\chi$PT is remarkably flat
that is perfectly consistent with the striking linear behaviour of the mode number over the same range
(cfr. fig. 3 of ref.~\cite{Giusti:2008vb}).
The last remark concerns the behaviour of the spectral density close to the threshold: here the NLO corrections coming from the O($a^2$) terms become very large. On the other hand, we don't expect our formula eq.~\ref{eq:rho_Q} to be valid close to the threshold. 
The reason is that close to the threshold
(i.e. $\lambda \rightarrow m$) at fixed lattice spacing $a$ and at fixed quark mass $m$ 
the relevant scale $\sqrt{\lambda^2-m^2}$, related to the valence polar mass, might become very small.
It is plausible then that for this scale
\begin{enumerate}
\item the GSM power counting $\sqrt{\lambda^2-m^2}\sim a\Lambda_{QCD}^2$ breaks down, and it is more appropriate to adopt the so called Aoki power counting (see ref.~\cite{Sharpe:2006ia});
\item the $p$-regime power counting, $M^2_\lambda L\gg 1$ (with $M_{\lambda}^2 = 2\Sigma \sqrt{\lambda^2 - m^2}/F^2$) breaks down and this mass scale enters the so-called $\epsilon$-regime. The FSE diverge when $\lambda \rightarrow m$~\cite{Necco:2011vx}. A possible explanation to this behaviour is indeed that
the zero-modes of the Goldstone bosons in the valence sector have to be treated in a 
non-perturbative manner as if the valence polar mass would be in the $\epsilon$-regime
(for example in refs.~\cite{Akemann:2010em,Splittorff:2011bj} both valence and 
sea quarks are considered in the $\epsilon$-regime).   
\end{enumerate}

\section{The spectral density close to the threshold}
A possible way to overcome the second limitation illustrated at the end of the previous section is to adopt the so-called \emph{mixed Chiral Effective Theory}~\cite{Bernardoni:2007hi}, where some masses obey the $p$-regime counting, and others are in the $\epsilon$-regime.\\
We first start with some considerations in the continuum. We take $N_s$ sea quarks with mass $m_s$ in the $p$-regime
and $N_v=1$ valence quarks in the $\epsilon$-regime, viz.
\be
m_s\sim O(p^2),\quad m_v \sim O(p^4), \quad 1/L,1/T \sim  O(p)\,. 
\label{eq:pow}
\ee
As an intermediate step of our calculation we introduce a $\theta$-term as follows
\be
\mathcal{L}_{2} = \frac{F^2}{4}
{\rm Tr}\left[\partial_\mu U\partial_\mu U^\dagger\right]-\frac{\Sigma}{2}{\rm Tr} \left[U_\theta^\dagger U(x)^\dagger\mathcal{M}+ \mathcal{M}^\dagger U(x) U_\theta\right]\,.
\ee
It is important to remark that the introduction of a $\theta$-term it is just
a computational tool, and this remark becomes more important when we extend the framework to W$\chi$PT. 
For simplicity we consider the replica formulation\footnote{The same conclusions are valid if we use the supersymmetric formulation.},
where
\be
\mathcal{M} = \mathcal{M}^{\dagger}={\rm diag} (\underbrace{m_s,m_s,\ldots,m_s}_{N_s}, 
\underbrace{m_v,m_v,\ldots,m_v}_{N_r})\,,\,\,\,\,\,\,\,\,\,\,\,\,\,\,\,\,\,\,U_\theta={\rm diag}(e^{\frac{i\theta}{N_s}\mathbbm{1}_s}, \mathbbm{1}_r) \,.
\label{eq:massterm}
\ee
While in the continuum it is not important how the $\theta$-term is introduced in the 
parametrization of the $U$-field~\cite{Bernardoni:2007hi,Bernardoni:2008ei},
we have decided to introduce the $\theta$-term only in the sea sector because it becomes
relevant when extending our calculation to finite lattice spacing.

It turns out~\cite{Bernardoni:2007hi} that among the Goldstone bosons there is one degree of freedom 
that becomes massless when we send the number of replicas to zero $N_r \rightarrow 0$. 
It is the Goldstone boson that is a singlet under the $SU(N_r)$ subgroup
of $SU(N_s+N_r)$. To overcome this problem one treats the zero-modes related to this degree of freedom, 
labelled by $\eta$, in a non-perturbative manner~\cite{Bernardoni:2007hi}. The $U$-field can then be parametrized 
as follows
\be
U(x) = \left( \begin{array}{cc}
{\rm e}^{-i\frac{\eta}{N_s}\mathbbm{1}_s} &  \\
 & \underbrace{{\rm e}^{i\frac{\eta}{N_r}\mathbbm{1}_r}U_0}_{\overline{U}_0}  \end{array} \right) {\rm e}^{\frac{2i\xi(x)}{F}}\,.
\ee
In the {\it valence} sector additionally to the zero-modes field $U_0$, needed because the valence quarks are in the
$\epsilon$-regime, we have the $\eta$-field that makes $\overline{U}_0 \in U(N_r)$. In the {\it sea} sector we just 
have the additional $\eta$-field and as a consequence the fluctuations $\xi$ do not contain the zero-modes 
of the $SU(N_r)$ generators nor of the $\eta$-field. We can now shift the $\theta$-term in the parametrization
of the $U$-field as follow
\be
U(x)U_\theta\rightarrow U(x).
\label{eq:U_para}
\ee
The periodicity in $\theta$ of the chiral Lagrangian allows us to write the partition function
in standard fashion
\be
\mathcal{Z}(\theta) = \sum_{\nu=-\infty}^{\nu=+\infty} e^{-i \nu \theta}\mathcal{Z}_\nu\,, \qquad 
\mathcal{Z}_\nu = \frac{1}{2 \pi} \int_0^{2 \pi} d \theta~ e^{i \nu \theta} \mathcal{Z}(\theta)\,.
\ee
By performing an exact integration over the constant field $\bar{\theta}=\theta-\eta$ one obtains 
~\cite{Bernardoni:2007hi,Bernardoni:2008ei}
\be
\mathcal{Z}_\nu \propto e^{\frac{-N_s \nu^2}{2 z_s}} \int_{U(N_r)} d \overline{U}_0 
\left( \det \overline{U}_0 \right)^\nu {\rm e}^{\frac{m_v \Sigma V}{2} 
{\rm Tr}\left[ \overline{U}_0^\dagger + \overline{U}_0 \right]}, 
\qquad z_s = m_s V \Sigma\,,
\label{eq:Znu}
\ee
from which one observes that the distribution of $\nu$ is Gaussian
and it is controlled by the sea quarks which are in the $p$-regime.
The computation of the spectral density is now straightforward and we obtain the NLO result
\be
\rho(\zeta) = \sum_{\nu=-\infty}^{\nu=+\infty} \rho_\nu(\zeta) \frac{\mathcal{Z}_\nu}{\mathcal{Z}}\,, \qquad 
\rho_\nu(\zeta) = \frac{\Sigma_{\rm eff}}{2}\zeta\left[J_\nu^2(\zeta) - J_{\nu +1}(\zeta)J_{\nu -1}(\zeta) \right]\,, 
\qquad
\zeta=\gamma \Sigma V \,,
\label{eq:rho_mixed}
\ee
where $\Sigma_{\rm eff}$ is a function of the sea quark mass $m_s$ and of the geometry of the $4$-d box. \footnote{See for instance \cite{Bernardoni:2010nf} for an explicit expression of $\Sigma_{\rm eff}$.}
The sum over $\nu$ can be easily done because we know the weight factor $\frac{\mathcal{Z}_\nu}{\mathcal{Z}}$
(cfr. eq.~\ref{eq:Znu}).
We observe that the shape of the spectral density is entirely described 'effectively' by the spectral 
density in the quenched theory, while the effects of the sea quarks are twofold. They change the absolute
normalization by introducing an $m_s$ dependence in $\Sigma_{\rm eff}$ and they control the distribution of $\nu$,
which, we remark, stays Gaussian only because the sea quarks are in the $p$-regime.

To extend this result to include the lattice spacing effects we need to discuss how to deal with the $\theta$-term
and the O($a$) already at the LO of the chiral Lagrangian
\be
\mathcal{L}_2 = \frac{F^2}{4}{\rm Tr}\left[\partial_\mu U\partial_\mu U^\dagger\right] -
\frac{\Sigma}{2}{\rm Tr}\left[U_\theta^\dagger U(x)^\dagger\mathcal{M}+ \mathcal{M}^\dagger U(x) U_\theta\right]-
\frac{\hat{a}F^2}{4}{\rm Tr}\left[U+U^\dagger\right]\,,
\ee
where the mass matrix is of the same form as in eq.~\ref{eq:massterm}, but with a twisted mass $(m_v+ i\mu_v\tau^3)$ in the valence sector.
It becomes evident now the importance of introducing a $\theta$-term only in the sea sector. 
Exactly as we have done for the continuum theory we redefine the $U$-field as in eq.~\ref{eq:U_para}
and the resulting chiral Lagrangian becomes
\be
\mathcal{L}_2 = \frac{F^2}{4}{\rm Tr}\left[\partial_\mu U\partial_\mu U^\dagger\right] -
\frac{\Sigma}{2}{\rm Tr}\left[\mathcal{M}U^\dagger+U\mathcal{M}^\dagger\right]-
\frac{\hat{a}F^2}{4}{\rm Tr}\left[U_\theta^\dagger U+U_\theta U^\dagger\right]\,.
\ee
The important question is to understand how to reabsorb the O($a$) term in a redefinition
of the mass term in presence of a $\theta$-term. One possible answer is based on a combination of introducing
the $\theta$ term only in the sea sector and an appropriate choice of power counting.
Once we have fixed the power counting for the masses as in eq.~\ref{eq:pow}, where we now have $m_P=\sqrt{m_v^2+\mu_v^2}\sim O(p^4)$,
we have to decide which power counting we want to adopt for the lattice spacing $a$. We can consider the following cases:
\begin{itemize}
\item $a\sim m_P\sim p^4$ (\emph{GSM-valence}): at NLO, the sea quarks are effectively as in the continuum;\vspace{-0.2cm}
\item $a\sim p^3$ (\emph{GSM$^*$-valence}~\cite{Shindler:2009ri,Bar:2008th}): at NLO, the sea quarks are effectively in the continuum;\vspace{-0.2cm}
\item $a\sim m_s\sim p^2$ (\emph{GSM-sea/Aoki-valence}): at NLO, lattice spacing corrections affect both the sea and the valence sector.\vspace{-0.2cm}
\end{itemize}
It is easy to see that in the first two power countings the O($a$) can be reabsorbed in the untwisted valence mass $m_v$,
because in the valence sector there is no $\theta$-term, due to the specific choice of $U_\theta$.  The O($a$) corrections in the sea sector are, on the other side, of higher order, leaving all the unwanted mixing between 
$\theta$-term and O($a$) as NNLO corrections. \\
Another important results of this mechanism is that in the GSM and GSM$^*$-valence power countings 
the distribution of $\nu$ is still Gaussian and the lattice spacing corrections appear at NNLO. 
This comes hardly as as a surprise
because in the mentioned power countings the sea quarks, that control the distribution of $\nu$, are effectively 
in the continuum. While it is still possible that discretization effects will suffer from the first limitation exposed at the end of section 3, 
we believe that this framework could be the appropriate one to describe numerical data obtained with quark masses in the $p$-regime (as usually is done with Wilson fermions).

\section{Conclusions}
We have computed the spectral density in the $p$-regime of W$\chi$PT at NLO~\cite{Necco:2011vx}. 
Our final result eq.~\ref{eq:rho_Q} gives a good description of the numerical 
data available~\cite{Giusti:2008vb}. Lattice artefacts at NLO are small and vanish when $m \rightarrow 0$
if we stay sufficiently distant from the threshold. 
Our formula can help both in estimating the safe region in the eigenvalue spectrum
to use for the extraction of $\Sigma$ and also to give an estimate of $W_8'$.\\
The theoretical understanding of the behaviour of the spectral density near the threshold is not complete and 
to fill this gap we have started a computation in the $\epsilon$/$p$-regime in PQW$\chi$PT. 
We have proposed a computational tool to deal with the simultaneous presence of a $\theta$-term and O($a$) corrections in the
chiral Lagrangian. \\
We acknowledge useful discussions with L.~Giusti and P.~Hern\'andez.

\bibliographystyle{h-elsevier}
\bibliography{sd_lat11}

\end{document}